\documentclass{osa-article}

\journal{oe}

\usepackage{graphicx}
\usepackage{xcolor}
\usepackage{siunitx}
\usepackage{hyperref}
\usepackage{svg}

\hypersetup{
    colorlinks=true,
    linkcolor=blue,
    citecolor=blue,
    filecolor=blue,
    urlcolor=blue
}

\begin{document}

\title{Electro-optic frequency comb Doppler thermometry}

\author{Sean M. Bresler,\authormark{1, 2, 4} Erin M. Adkins,\authormark{3} Stephen P. Eckel,\authormark{1} Tobias K. Herman,\authormark{1} David A. Long,\authormark{1} Benjamin J. Reschovsky\authormark{1} and Daniel S. Barker\authormark{1, 5}}

\address{\authormark{1}Physical Measurement Laboratory, National Institute of Standards and Technology, Gaithersburg, MD 20899, USA\\
\authormark{2}Department of Chemistry, University of Maryland, College Park, MD 20742, USA\\
\authormark{3}Material Measurement Laboratory, National Institute of Standards and Technology, Gaithersburg, MD 20899, USA\\
\authormark{4}sean.bresler@nist.gov\\
\authormark{5}daniel.barker@nist.gov}


\begin{abstract*}
We demonstrate a Doppler thermometer based on direct optical frequency comb spectroscopy of an \(^{85}\)Rb vapor with a chirped electro-optic frequency comb (EOFC).
The direct EOFC Doppler thermometer is accurate to within its approximately \(1~\si{\kelvin}\) statistical uncertainty.
We experimentally compare direct EOFC spectroscopy with conventional Doppler spectroscopy using a single-frequency, step-scanned laser probe.
Our results show that direct EOFC spectroscopy mitigates transit-induced optical pumping distortion of the atomic lineshape, which is the dominant systematic temperature shift in alkali atom Doppler thermometry.
Optical Bloch equation simulations of conventional and direct EOFC Doppler spectroscopy confirm that EOFC spectroscopy can use higher optical power to reduce statistical noise without optical pumping distortion.
Our results indicate that EOFC Doppler thermometry is a promising approach to realizing a primary thermometer with size and measurement rate sufficient for applications including pharmaceutical manufacturing and nuclear waste monitoring.
\end{abstract*}

\section{\label{sec:intro}Introduction}

Motivated by the 2019 SI redefinition, there is a growing effort to develop portable, stable, manufacturable thermometers based on primary measurement methods~\cite{hendricks2020, dedyulin2022}.
Approaches currently under investigation include Johnson noise thermometry~\cite{bramley2016}, quantum-correlation thermometry~\cite{Purdy2017}, and Doppler broadening thermometry~\cite{Agnew2025}.
All approaches aim to produce a thermometer that has fast enough measurement rate, small enough size, and low enough uncertainty for industrial and metrological applications.
These three goals -- fast, small, accurate -- are typically in conflict.
For example, recently demonstrated quantum-correlation thermometers are small and accurate, but must average for approximately \(100~\si{\second}\) to reach approximately \(10~\si{\kelvin}\) uncertainty near \(300~\si{\kelvin}\)~\cite{Purdy2017}.
Doppler thermometry can be fast and accurate, but achieving sufficient absorption often requires a large vapor cell.
In particular, Doppler thermometry using molecular gases typically requires vapor cells that are 10~cm long or longer~\cite{Daussy2007, Casa2008,Gotti2018,Galzerano2020}.

Realizing a Doppler thermometer with a size of a few millimeters, comparable to platinum resistance thermometers, will likely require using an atomic gas.
Because atomic gases can exhibit strong optical absorption at low vapor pressure~\cite{truong2011, truong2015a}, an atomic Doppler thermometer could retain sufficiently large absorption using a millimeter-scale vapor cell.
Laboratory-size Doppler thermometers based on rubidium, cesium, and mercury have all demonstrated uncertainty sufficient for many applications~\cite{truong2011, truong2015, gravina2024, Agnew2025}.
Promisingly, millimeter-scale microfabricated vapor cells of the alkalis rubidium and cesium with high absorption have been demonstrated, but, to our knowledge, have yet to be used for Doppler thermometry~\cite{Liew2004,Kitching2018,Bopp2020}.

One particular issue with an alkali-based Doppler thermometer is transit-induced lineshape distortion~\cite{stace2010, stace2012}.
The transit-induced distortion arises from optical pumping into states that are transparent to the spectroscopy laser.
Such transparent states occur in alkali atoms because the spectroscopy laser is only resonant with transitions from one of the atom's hyperfine ground states.
As an atom passes through the laser beam, the amount of optical pumping it experiences is determined by the laser intensity and its transit time in the beam.
When convolved with the Maxwell-Boltzmann velocity distribution, the optical pumping distorts the absorption lineshape away from the expected Voigt profile~\cite{stace2010, stace2012}.

The transit-induced distortion causes a systematic underestimate of the atomic vapor temperature, which is the dominant systematic shift even for modest laser intensity~\cite{siddons2008, stace2012, truong2015, Agnew2025}.
Beyond-Voigt fitting models that correct the temperature shift are perturbative and therefore only accurate at low intensity~\cite{stace2012, truong2015}.
Thus, another competition arises in alkali-based Doppler thermometers: more laser intensity, which allows for faster averaging in the limit of photon shot noise, induces a larger systematic uncertainty in the measured temperature, compromising accuracy.
Finding an interrogation strategy that allows for faster averaging without comprising accuracy therefore becomes an important goal for making a deployable Doppler thermometer.


Here, we demonstrate atomic Doppler thermometry using a chirped electro-optic frequency comb (EOFC).
In a chirped EOFC, an electro-optic modulator (EOM) is weakly driven with a repetitively frequency-chirped sinusoid~\cite{Long2016, Long2019}.
In contrast to overdriven EOFCs~\cite{Parriaux2020}, the EOM in a chirped EOFC imprints \(1\)st-order modulation sidebands on the spectroscopy beam, while keeping higher-order sidebands well below detection limits.
Averaged over many frequency chirps, the EOM output forms an optical frequency comb with the comb tooth spacing given by the frequency chirp repetition rate.
The comb teeth provide a precise frequency scale for fitting and temperature extraction (see Sec.~\ref{sec:apparatus}).

The chirped EOFC significantly reduces transit-induced distortion when it directly probes the atomic vapor.
Because the EOFC repetition rate exceeds the natural decay rate, atoms do not equilibrate with the spectroscopy beam as they pass through it.
Optical Bloch equation simulations of the optical pumping process indicate that transit-induced distortion only occurs in equilibrium (see Sec.~\ref{sec:sim}).
As a result, direct EOFC Doppler thermometry does not suffer from transit-induced distortion.
We confirm the absence of transit-induced distortion by experimentally comparing direct EOFC Doppler thermometry to conventional Doppler thermometry on the D\(_2\) line of \(^{85}\)Rb in Sec.~\ref{sec:res}.
Our results show that a direct EOFC Doppler thermometer can use high optical power to reach large signal-to-noise ratios without systematic shifts, allowing low uncertainty temperature measurement at rates faster than a conventional Doppler thermometer.

\section{\label{sec:apparatus}Apparatus}

\begin{figure}
\centering\includesvg[width=\columnwidth]{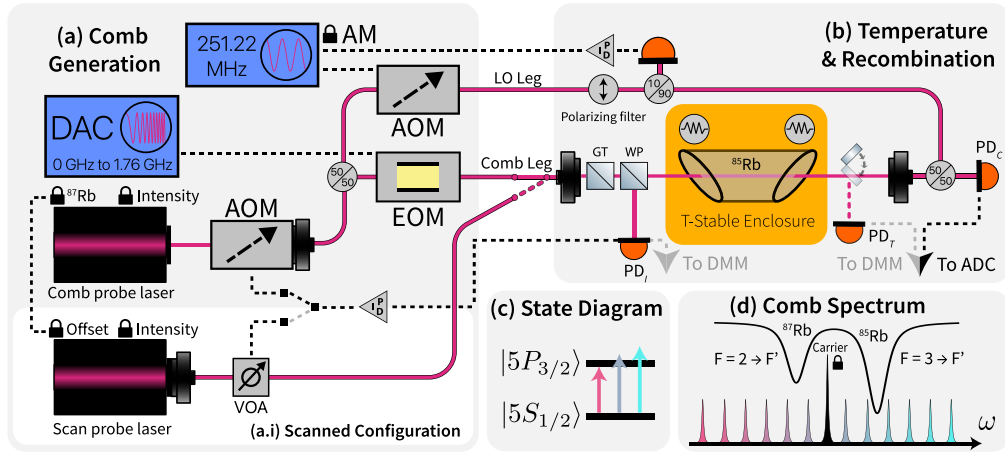}
\caption{\label{fig:apparatus}
    Schematic of the apparatus and measurement scheme: (a) Electro-optic frequency comb (EOFC) generation system and scan probe laser (shown in (a.i)), (b) temperature measurement and EOFC recombination setup, (c) state diagram for the rubidium D\(_2\) line, (d) EOFC spectrum with the relevant rubidium hyperfine transitions.
    Abbreviations -- ADC: analog-to-digital converter, AM: amplitude modulation, AOM: acousto-optic modulator, DAC: digital-to-analog converter, DMM: digital multimeter, EOM: electro-optic modulator, GT: Glan-Thompson polarizer, LO: local oscillator, PD: photodiode, PID: proportional-integral-differential controller, VOA: variable optical attenuator, WP: Wollaston prism.
    The Glan-Thompson polarizer and Wollaston prism are depicted as beamsplitting cubes.
    }
\end{figure}

We test the chirped EOFC spectroscopy approach to atomic Doppler thermometry by measuring the temperature of an \(^{85}\)Rb vapor.
Figure~\ref{fig:apparatus} shows our Doppler spectroscopy apparatus.
Light for chirped EOFC spectroscopy and conventional stepped scan spectroscopy is generated by two \(780~\si{\nano\meter}\) lasers, which we designate as the comb probe laser and the scan probe laser, respectively.
Optical bandpass filters with a \(3~\si{\nano\meter}\) full-width at half-maximum block amplified spontaneous emission from both lasers.
The comb probe laser is frequency stabilized using frequency-modulated, sub-Doppler spectroscopy of the rubidium \(|5\, ^{2}S_{1/2}\rangle\rightarrow |5\, ^{2}P_{3/2}\rangle\) transition, see Fig.~\ref{fig:apparatus}(c).
The scan probe laser is offset locked to the comb probe laser using an optical phase-locked loop (OPLL).

Our EOFC generation system, shown in Fig.~\ref{fig:apparatus}(a), is similar to those described in Refs.~\cite{Long2016,Long2019,Long2021,Reschovsky2022}.
In this system, we pass light from the comb probe laser through an acousto-optic modulator (AOM), for intensity stabilization, and then fiber-couple it to the EOFC generation system.
A fiber beamsplitter divides the light between a comb leg and a local oscillator (LO) leg.
A digital-to-analog converter (DAC) weakly drives the EOM in the comb leg with a repeating, linear frequency chirp, so that the instantaneous detuning of the \(\pm 1\)st-order sidebands relative to the carrier at time \(t\) is
\begin{equation}
    \label{eq:comb_det}
    \Delta(t)= \pm\Big(\Delta_0 + \Delta_m (t\,\mathrm{mod}\,t_{\rm rep})/t_{\rm rep}\Big),
\end{equation}
where \(\Delta_0\) is the minimum angular frequency of the DAC output sine wave, \(\Delta_m\) is the angular frequency amplitude of the frequency chirp, and \(1/t_{\rm rep}\) is the chirp repetition rate.
Averaged over multiple chirps, Eq.~\eqref{eq:comb_det} converts the light output from the EOM into an optical frequency comb with span \(2\Delta_m\) and tooth spacing \(2\pi/t_{\rm rep}\).
The comb teeth arise from first-order phase modulation and are therefore all low power.
The averaging time can be lengthened to minimize frequency uncertainty or shortened to maximize temporal resolution.
We choose \(\Delta_0=2\pi\times 10.56~\si{\mega\hertz}\), \(\Delta_m=2\pi\times 1.7~\si{\giga\hertz}\), \(1/t_{\rm rep}=10.56~\si{\mega\hertz}\), and an averaging time of \(0.5~\si{\second}\).
The data clock rate for the DAC is \(5.407\times 10^9\)~samples~per~second, which is greater than \((\Delta_0+\Delta_m)/\pi\).
Importantly, \(2\pi/t_{\rm rep}\approx 6.6\times 10^7~\si{\per\second}\) is greater than the natural decay rate \(\Gamma \approx 3.811\times 10^7~\si{\per\second}\) of the rubidium \(|5\, ^{2}S_{1/2}\rangle\rightarrow |5\, ^{2}P_{3/2}\rangle\) transition~\cite{Steck2008}.

To read out the EOFC spectrum, we perform a heterodyne measurement by interfering the EOFC light with light from the LO leg on a fast photodiode (PD\(_C\) in Fig.~\ref{fig:apparatus}(b)).
The fast PD converts the optical beats between the LO and the EOFC into an electronic signal that is recorded by a fast analog-to-digital converter (ADC) at a rate of \(3\times 10^9\)~samples~per~second.
The analog bandwidth of the ADC is less than its Nyquist frequency, so it does not detect high frequency artifacts produced by the DAC.
Light in the LO leg passes through an AOM that frequency shifts it by 251.22~\si{\mega\hertz}.
The LO-leg AOM drive frequency is incommensurate with \(1/t_{\rm rep}\), which ensures that electronic signals due to \(+1\)st-order and \(-1\)st-order comb teeth occur at different frequencies. 
Proportional-integral-differential (PID) feedback to the RF drive amplitude of the LO-leg AOM stabilizes the LO light intensity.

To perform direct EOFC Doppler thermometry, we measure the transmission of the chirped EOFC light through a Brewster-cut rubidium vapor cell.
The vapor cell contains isotopically purified \(^{85}\)Rb, is \(75~\si{\milli\meter}\) long, and is enclosed in a temperature-stabilized aluminum mount.
Two calibrated industrial platinum resistance thermometers (PRTs) record the mount's temperature, which is stabilized to \(298.150(15)~\si{\kelvin}\) by resistive heaters.
(Here, and throughout the paper, parenthetical quantities represent the standard uncertainty).
The uncertainty in the vapor cell temperature is given by the \(15~\si{\milli\kelvin}\) gradient across the aluminum mount, since the 
\(3~\si{\milli\kelvin}\) PRT calibration uncertainty is negligible.
The EOFC light is collimated to a \(1/e^2\) radius of \(0.8~\si{\milli\meter}\), polarized by a Glan-Thompson polarizer, and then split \(50{:}50\) by a Wollaston prism.
The EOFC light from one output of the prism forms a probe beam that transmits the vapor cell and is subsequently fiber-coupled to PD\(_C\), where it interferes with the LO light.
The EOFC light from the other output of the prism is detected by a PD in photovoltaic mode (PD\(_{I}\) in Fig.~\ref{fig:apparatus}(b)), which stabilizes the EOFC light intensity via PID feedback to the AOM at the comb probe laser output, see Fig.~\ref{fig:apparatus}(a).

We compute the transmission of the \(^{85}\)Rb vapor by recording EOFC power spectra in three conditions.
First, we lock the comb probe laser to the \(|5\, ^{2}S_{1/2}, F=2\rangle\rightarrow |5\, ^{2}P_{3/2}, F'=3\rangle\) transition of \(^{87}\)Rb via saturated absorption spectroscopy (where \(F\) and \(F'\) are the total angular momentum quantum numbers for the ground and excited states, respectively).
The saturated absorption spectroscopy setup is not shown in Fig.~\ref{fig:apparatus}.
The \(^{87}\)Rb \(|5\, ^{2}S_{1/2}, F=2\rangle\rightarrow |5\, ^{2}P_{3/2}, F'=3\rangle\) lockpoint puts the EOFC light on resonance with the \(^{85}\)Rb transitions from \(|5\, ^{2}S_{1/2}, F=3\rangle\) to \(|5\, ^{2}P_{3/2}\rangle\) (see Fig.~\ref{fig:apparatus}(c) and Fig.~\ref{fig:apparatus}(d)), so the recorded spectra measure the power transmitted through the vapor near resonance.
Second, we lock the comb probe laser to the \(|5\, ^{2}S_{1/2}, F=1\rangle\rightarrow |5\, ^{2}P_{3/2}, F'=2\rangle\) transition of \(^{87}\)Rb.
The \(^{87}\)Rb \(|5\, ^{2}S_{1/2}, F=1\rangle\rightarrow |5\, ^{2}P_{3/2}, F'=2\rangle\) lockpoint puts the EOFC light far from resonance with any \(^{85}\)Rb transitions, so the recorded spectra measure the power transmitted by the vapor off resonance, which we use as a proxy for the power incident to the vapor.
Third, we block both the EOFC and LO light from PD\(_C\), so the recorded spectra measure the dark current background of PD\(_C\).
We subtract the dark spectra from the resonant and non-resonant spectra.
The \(^{85}\)Rb vapor transmission \(\mathcal{T}\) is then given by ratio of the background subtracted resonant and non-resonant spectra.

To measure the vapor transmission with conventional stepped scans~\cite{truong2011, truong2015}, we replace the EOFC light with light from the scan probe laser, see Fig.~\ref{fig:apparatus}(a) and Fig.~\ref{fig:apparatus}(a.i).
We lock the comb probe laser to the \(|5\, ^{2}S_{1/2}, F=1\rangle\rightarrow |5\, ^{2}P_{3/2}, F'=1\rangle\) transition of \(^{87}\)Rb and step the scan probe laser across the transitions from the \(^{85}\)Rb \(|5\, ^{2}S_{1/2}, F=3\rangle\) ground state by varying its frequency offset from the comb probe laser.
Photodiodes synchronously measure the power incident to and transmitted by the vapor cell (PD\(_{I}\) and PD\(_{T}\) in Fig.~\ref{fig:apparatus}(b), respectively).
We operate our PDs in photovoltaic mode to minimize dark current.
The power incident to the vapor cell is also stabilized by PD\(_{I}\) via PID feedback to a voltage-variable optical attenuator, see Fig.~\ref{fig:apparatus}(a.i).
The current output of both PDs is converted to voltage using high-linearity transimpedance amplifiers and digitized by \(6.5\)-digit multimeters.
To correct for PD dark current, we measure the signal from both PD\(_I\) and PD\(_T\) with the incident laser beam blocked.
We compute \(\mathcal{T}\) for a stepped scan from the ratio of the dark-current-subtracted PD signals.

\section{\label{sec:sim}Simulation}

To understand the transit-induced distortion, we simulate the optical pumping as atoms transit the spectroscopy laser beam using the optical Bloch equations (OBEs)~\cite{Ungar1989, Eckel2022}.
We follow the approach of prior studies of transit-induced distortion in Doppler thermometry~\cite{stace2010, stace2012}, where the optical pumping was investigated using a simplified \(\Lambda\)-type level structure.
We label the three energy levels \(|g\rangle\), \(|d\rangle\), and \(|e\rangle\).
The excited state \(|e\rangle\) decays at the natural decay rate \(\Gamma\) of the rubidium \(|5\, ^{2}S_{1/2}\rangle\rightarrow |5\, ^{2}P_{3/2}\rangle\) transition and the branching ratio between states \(|g\rangle\) and \(|d\rangle\) is \(1/2\).
The spectroscopy laser with intensity \(I_{ge}\) couples \(|g\rangle\) to \(|e\rangle\) at a Rabi frequency \(\Omega = \Gamma\sqrt{s/2}\), where \(s=I_{ge}/I_{\rm sat}\) is the saturation parameter and \(I_{\rm sat}\) is the saturation intensity of the \(|g\rangle\rightarrow|e\rangle\) transition.
We model the spectroscopy laser beam as a flat-top beam with a radius that matches the \(1/e^2\) radius of the laser beam in our experiment (see Sec.~\ref{sec:apparatus}).

We compute the distorted lineshape using the instantaneous scattering rate on the \(|g\rangle\rightarrow|e\rangle\) transition, which is given by the imaginary component of the off-diagonal element of the density matrix \(\rho_{ge}\). 
For stepped scan simulations, we fix the spectroscopy laser detuning \(\Delta\) and numerically integrate the OBEs for a stationary atom at the center of the laser beam to find \(\mathrm{Im}[\rho_{ge}(s, \Delta, t)]\) at integration time \(t\).
To generate the lineshape \(\mathrm{Im}[\rho_{ge}(s, \Delta)]\), we repeat the simulation at various \(\Delta\) and average over the distribution of dwell times of an atom in the laser beam.
We calculate the dwell time distribution function \(\mathcal{H}(t)\) from the laser beam radius and the Maxwell-Boltzmann-distributed atomic velocity perpendicular to the laser beam following the method of Ref.~\cite{Harris2006}. 
The lineshape is then given by
\begin{equation}
    \label{eq:trans_dist}
    \mathrm{Im}[\rho_{ge}(s, \Delta)] = A \int_0^{t_{\rm max}} \mathrm{Im}[\rho_{ge}(s, \Delta, t)] \mathcal{H}(t) dt,
\end{equation}
where \(A\) is a normalization constant and \(t_{\rm max}\) is the maximum OBE integration time.
We choose \(t_{\rm max}\) such that \(\mathcal{H}(t_{\rm max})\approx 0\).
When \(s=0\), \(\mathrm{Im}[\rho_{ge}(s=0, \Delta)] = A/(1+4\Delta^2/\Gamma^2)\) is the expected Lorentzian lineshape.

For EOFC simulations, the laser detuning \(\Delta(t)=\Delta_0+\Delta_m (t\,\mathrm{mod}\,t_{\rm rep})/t_{\rm rep}\).
We take \(\Delta_0\), \(\Delta_m\), and \(t_{\rm rep}\) matching the experimental conditions (see Sec.~\ref{sec:apparatus}).
Numerical integration of the OBEs for a stationary atom at the center of the laser beam thus yields \(\mathrm{Im}[\rho_{ge}(s, t)]\), which includes information at all \(\Delta\).
We then use the discrete Fourier transform to construct the power spectrum of \(\mathrm{Im}[\rho_{ge}(s, t)]\) while averaging over \(\mathcal{H}(t)\).
The resulting power spectrum is
\begin{equation}
    \label{eq:pow_spec}
    \mathrm{Im}[\rho_{ge}(s, \omega)] = A\, \bigg|\sum_{n=0}^{n=t_{\rm max}/\delta t}\mathcal{H}(n\delta t)\mathrm{Im}[\rho_{ge}(s, n\delta t)]e^{i\omega n\delta t}\bigg|^2,
\end{equation}
where \(n\) is an integer, \(i\) is the imaginary unit, \(A\) is a normalization constant, and \(\delta t\) is the sampling interval of the OBE simulation.
Because the laser detuning is periodic, \(\mathrm{Im}[\rho_{ge}(s, \omega)]\) exhibits sharp peaks when \(\omega\) is an integer multiple of \(1/t_{\rm rep}\), which we extract with a peak finding algorithm.
The simulated EOFC lineshape is then \(\mathrm{Im}[\rho_{ge}(s, \Delta)] = \mathrm{Im}[\rho_{ge}(s, \omega)]|_{\omega=n/t_{\rm rep}}\).

\begin{figure}
\centering\includesvg[width=\textwidth]{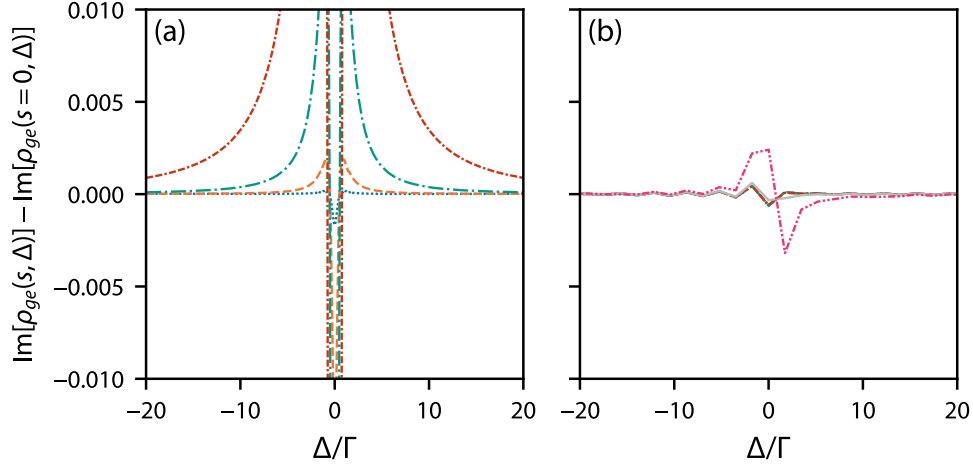}
\caption{\label{fig:sim}
    Difference between distorted and Lorentzian lineshapes for (a) stepped scan spectroscopy and (b) direct EOFC spectroscopy.
    The effect of transit-induced distortion is shown for \(s=10^{-4}\) (blue, dotted), \(s=10^{-3}\) (orange, dashed), \(s=10^{-2}\) (teal, dash-dotted), \(s=10^{-1}\) (red, dash-dash-dotted), \(s=1\) (gray, solid), and \(s=10\) (magenta, dash-dot-dotted).
    }
\end{figure}

We determine the size of the transit-induced distortion by taking the difference between \(\mathrm{Im}[\rho_{ge}(s, \Delta)]\) and \(\mathrm{Im}[\rho_{ge}(s=0, \Delta)]\).
Figure~\ref{fig:sim} shows \(\mathrm{Im}[\rho_{ge}(s, \Delta)]-\mathrm{Im}[\rho_{ge}(s=0, \Delta)]\) over a range of \(s\) for both stepped scan and EOFC spectroscopy.
The stepped scan simulations in Fig.~\ref{fig:sim}(a) exhibit substantial transit broadening and distortion even at saturation parameters below \(s = 10^{-3}\).
When convolved with the Gaussian velocity distribution, the natural lineshape distortion perturbs the transmission spectrum away from the pure Voigt model.
The transit-induced distortion gives rise to an oscillation in the residuals of Voigt model fits to distorted transmission spectra and a systematic shift in fitted temperature, see Sec.~\ref{sec:res}.
The EOFC simulations in Fig.~\ref{fig:sim}(b) do not display substantial distortion until the saturation parameter is greater than \(s = 1\).
A direct EOFC Doppler thermometer can use a spectroscopy laser beam with \(s\ge 10^{-2}\) without needing to mitigate systematic temperature shifts via beyond-Voigt lineshape models~\cite{stace2010, stace2012}.

\section{\label{sec:res}Experimental Results}

\begin{figure}[t]
\centering\includesvg[width=\textwidth]{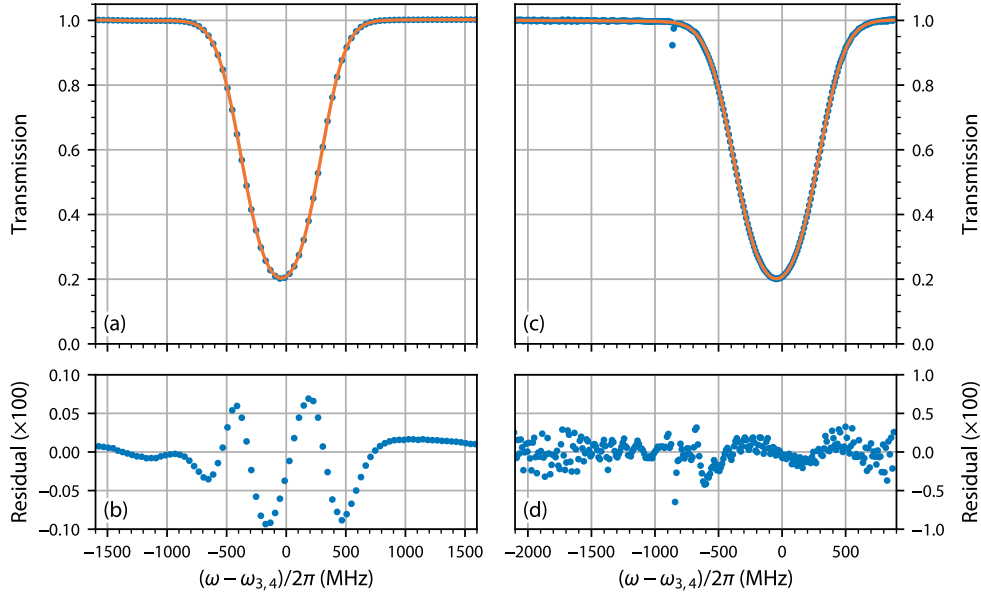}
\caption{\label{fig:example}
    Example stepped scan and direct EOFC Doppler spectra.
    (a) Stepped scan transmission spectrum at \(s\approx 0.1\) (blue points) with Voigt model fit (orange line).
    (b) Residuals for the fit and spectrum in (a).
    (c) Direct EOFC transmission spectrum at \(s\approx 0.4\) (blue points) with Voigt model fit (orange line).
    (d) Residuals for the fit and spectrum in (c).
    The residuals in (b) and (d) are plotted on different vertical scales.
    The feature in (b) near \((\omega-\omega_{3,4})/2\pi\approx -1200~\si{\mega\hertz}\) arises due to a small relative abundance of \(^{87}\)Rb in the vapor cell (see text).
    }
\end{figure}

We confirm the OBE simulation results of Sec.~\ref{sec:sim} by measuring the temperature of the \(^{85}\)Rb vapor with both direct EOFC and stepped scan Doppler spectroscopy.
In both approaches, we acquire transmission spectra and fit to a model based on the Beer-Lambert law for the rubidium \(|5\, ^{2}S_{1/2}\rangle\rightarrow |5\, ^{2}P_{3/2}\rangle\) transition~\cite{siddons2008, truong2011}
\begin{equation}
    \label{eq:voigt}
    \mathcal{T}(\omega) = \Big(a_0+a_1(\omega-\omega_{3,4})+a_2(\omega-\omega_{3,4})^2\Big)\,e^{-\sum_{j}\mathcal{A}_j\sum_{F}\sum_{F'}S_{F,F'}V(\omega-\omega_{F,F'}, \delta_D)},
\end{equation}
where \(\omega_{F,F'}\) is the resonance frequency of the \(|5\, ^{2}S_{1/2}, F\rangle\rightarrow |5\, ^{2}P_{3/2}, F'\rangle\) transition, \(j=85,87\) indexes the rubidium isotope, \(\mathcal{A}_j\) is the absorption prefactor for isotope \(j\), \(S_{F,F'}\) is the relative transition strength, and \(V(\omega-\omega_{F,F'}, \delta_D)\) is the Voigt profile for the \(|5\, ^{2}S_{1/2}, F\rangle\rightarrow |5\, ^{2}P_{3/2}, F'\rangle\) transition with \(1/e\) Doppler width \(\delta_D\).
The coefficients \(a_0\), \(a_1\), and \(a_2\) account for a varying transmission baseline due to large free-spectral-range parasitic optical etalons~\cite{truong2011,truong2015}.
In our fits, we fix the Lorentzian width of \(V(\omega-\omega_{F,F'}, \delta_D)\) to the natural decay rate \(\Gamma\) of the rubidium \(|5\, ^{2}S_{1/2}\rangle\rightarrow |5\, ^{2}P_{3/2}\rangle\) transition.
After fitting, the temperature is given by
\begin{equation}
    T = \frac{m c^2}{2 k_B \omega_0^2}\delta_D^2,
\end{equation}
where \(m\) is the mass of a rubidium atom, \(\omega_0\) is the resonance frequency of the \(^{85}\)Rb \(|5\, ^{2}S_{1/2}\rangle\rightarrow |5\, ^{2}P_{3/2}\rangle\) transition, \(c\) is the speed of light, and \(k_B\) is the Boltzmann constant.

We show example Doppler spectra with fits to Eq.~\eqref{eq:voigt} in Fig.~\ref{fig:example}.
The stepped scan spectrum in Fig.~\ref{fig:example}(a) has a probe beam saturation parameter \(s=I/I_{\rm sat}\approx 0.1\), where \(I_{\rm sat}=1.669~\si[per-mode=symbol]{\milli\watt\per\square\centi\meter}\) is the minimum saturation intensity of the \(|5\, ^{2}S_{1/2}\rangle\rightarrow |5\, ^{2}P_{3/2}\rangle\) transition~\cite{Steck2008}.
At such a high saturation parameter, optical pumping as atoms transit the probe beam distorts the natural atomic lineshape away from the pure Lorentzian that is assumed by the Voigt model of Eq.~\eqref{eq:voigt}.
The transit-induced distortion leads to a characteristic oscillation in the fit residuals (defined as the data minus the best fit, see Fig.~\ref{fig:example}(b)) and causes the fit temperature \(T\) to be systematically lower than the actual temperature of the vapor~\cite{truong2015}.
The high signal-to-noise ratio of stepped scans also allows us to detect a small, approximately \(0.2~\si{\percent}\) relative abundance of \(^{87}\)Rb in our vapor cell, which gives rise to the feature in Fig.~\ref{fig:example}(b) near \((\omega-\omega_{3,4})/2\pi\approx -1200~\si{\mega\hertz}\).
For the direct EOFC spectrum in Fig.~\ref{fig:example}(c), the first-order electro-optic sidebands have instantaneous saturation parameter \(s\approx 0.4\).
The outliers near \((\omega-\omega_{3,4})/2\pi\approx -800~\si{\mega\hertz}\) are due to the EOFC carrier and are not included in the fit to Eq.~\eqref{eq:voigt}.
The presence of transit-induced distortion is not apparent in the residuals of the fit to the EOFC spectrum (see Fig.~\ref{fig:example}(d)).
Transit-induced distortion systematically shifts the fit temperature even when it is not resolved in the residuals, which we have verified by simulating noisy, distorted spectra with the models of Refs.~\cite{stace2010, stace2012} and fitting to Eq.~\eqref{eq:voigt}.
We also note that the EOFC spectra do not resolve the absorption due to the small \(^{87}\)Rb abundance.

\begin{figure}[t]
\centering\includesvg[width=4.5 in]{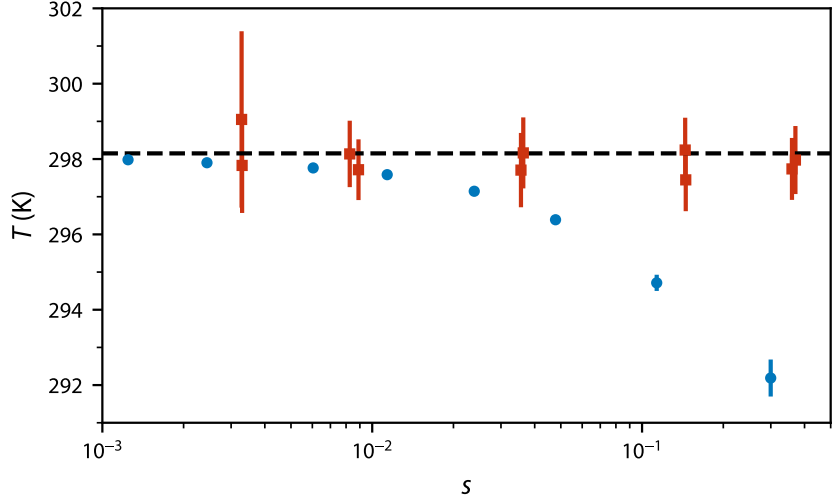}
\caption{\label{fig:temp}
    Fitted Doppler temperature as a function of probe saturation parameter.
    Blue circles and red squares show the temperature determined from stepped scan spectra and direct EOFC spectra, respectively.
    The horizontal dashed line denotes the temperature of the vapor cell reported by the two PRTs.
    Error bars represent the statistical standard uncertainty.
    Many error bars are smaller than the data points.
    The uncertainty in the vapor cell temperature is less than the width of the horizontal dashed line.
    }
\end{figure}

We study the relative importance of transit-induced distortion in direct EOFC and stepped scan thermometry by acquiring Doppler spectra at distinct probe intensities.
Figure~\ref{fig:temp} shows the temperature determined by a fit to Eq.~\eqref{eq:voigt} as a function of \(s\).
The temperature extracted from stepped scan spectra systematically shifts by \(6~\si{\kelvin}\) as \(s\) increases from approximately \(10^{-3}\) to approximately \(3\times10^{-1}\).
Accurate stepped scan thermometry requires either very low intensity or beyond-Voigt lineshape modeling~\cite{stace2010, stace2012, truong2015}.
In the photon-shot-noise limit, stepped scans therefore face a fundamental tradeoff between measurement single-to-noise ratio and systematic correction complexity, which may limit the rate at which stepped scans average down temperature uncertainty.
The temperature reported by direct EOFC spectra does not exhibit a systematic shift as \(s\) increases.
Direct EOFC thermometry can rely on Voigt models to much higher instantaneous intensities because direct EOFC spectroscopy divides the incident intensity across many comb teeth.
If the EOFC spectra can be measured near the photon-shot-noise limit, then direct EOFC thermometry should reach a target statistical uncertainty roughly \(\sqrt{\Delta_m t_{\rm rep}/2\pi}\) times faster than stepped scan thermometry.

The main sources of uncertainty for the stepped scan temperature measurements are statistical uncertainty, uncertainty in the hyperfine splitting of the \(^{85}\)Rb \(|5\, ^{2}P_{3/2}\rangle\) state, and the uncorrected transit-induced distortion shift.
At the lowest saturation \(s\approx 1.3\times 10^{-3}\), the statistical uncertainty of the stepped scan measurements is \(u_r(T) = 48~\si{\milli\kelvin}\).
Our temperature measurements use \(\omega_{F,F'}\) determined from a weighted average of the hyperfine constant measurements of Refs.~\cite{arimondo1977, Barwood1991, Rapol2003}.
We estimate the hyperfine splitting uncertainty by fitting Eq.~\eqref{eq:voigt} to simulated spectra generated with varying hyperfine constants and find \(u_{hf}(T) = 42~\si{\milli\kelvin}\).
We also use fits to simulated spectra to estimate that uncertainties due to magnetic fields, uncertainty in \(\Gamma\), and laser frequency noise are all small compared to \(u_r(T)\) and \(u_{hf}(T)\).
At \(s\approx 1.3\times 10^{-3}\), the stepped scan temperature is \(297.98(6)~\si{\kelvin}\), which is within \(170~\si{\milli\kelvin}\) of the temperature reported by the PRTs.
We have not corrected the temperature for the transit-induced distortion, because the approach of Refs.~\cite{stace2010, stace2012} has yet to be adapted to transitions with unresolved hyperfine structure, like \(|5\, ^{2}S_{1/2}, F=3\rangle\rightarrow |5\, ^{2}P_{3/2}, F'\rangle\). 
Even for transitions with resolved hyperfine structure (where Refs.~\cite{stace2010, stace2012} can be applied), the transit-induced distortion can be significant for \(s\approx 10^{-3}\)~\cite{truong2015}, so we believe the stepped scan temperature is in reasonable agreement with the PRTs.

The main sources of uncertainty for the direct EOFC temperature measurements are statistical uncertainty and uncertainty due to the \(^{87}\)Rb in the vapor cell.
For \(s\gtrsim 10^{-2}\), the statistical uncertainty of the direct EOFC measurements is \(u_r(T) \le 1~\si{\kelvin}\).
Because the comb probe laser is locked to the \(^{87}\)Rb \(|5\, ^{2}S_{1/2}, F=1\rangle\rightarrow |5\, ^{2}P_{3/2}, F'=2\rangle\) transition for off-resonance EOFC power measurements and \(^{87}\)Rb is approximately \(0.2~\si{\percent}\) abundant in the vapor cell, the measured non-resonant EOFC power spectra are impacted by \(^{87}\)Rb absorption.
The \(^{87}\)Rb absorption perturbs the transmission baseline away from the polynomial assumed in Eq.~\eqref{eq:voigt}.
We estimate the resulting systematic temperature shift \(\delta T\) by fitting Eq.~\eqref{eq:voigt} to simulated spectra with the baseline distorted by the \(0.2~\si{\percent}\) \(^{87}\)Rb abundance.
We find \(\delta T = -400(50)~\si{\milli\kelvin}\), where the uncertainty \(u_{\delta T}(T)\) arises from uncertainty in the \(^{87}\)Rb abundance and is negligible compared to \(u_r(T)\).
The direct EOFC temperature data in Fig.~\ref{fig:temp} are corrected for \(\delta T\) and agree with the temperature reported by the PRTs to within \(u_r(T)\).

The temperature uncertainty of the EOFC measurements is larger than the uncertainty of the stepped scan measurements by a factor of approximately \(10\).
We believe that our ability to reduce the direct EOFC thermometry \(u_r(T)\) with averaging is currently limited by long-term drifts in the EOFC power spectrum because the resonant and non-resonant EOFC power spectra are measured sequentially.
The direct EOFC transmission spectra are thus susceptible to baseline perturbation due to effects such as variations in DAC output amplitude, polarization rotations in optical fibers, and phase drifts between the two EOFC generation system legs.
In the future, such technical limitations could be overcome by implementing simultaneous readout of incident and transmitted spectra and/or phase-stabilization methods.
Direct EOFC thermometry could then fully realize the gains in averaging speed implied by EOFC spectroscopy's immunity to transit-induced distortion.

\section{\label{sec:con}Conclusion}

We experimentally and theoretically investigate the application of chirped electro-optic frequency combs to atomic Doppler thermometry.
Optical Bloch equation simulations of conventional stepped scan spectroscopy and direct EOFC spectroscopy indicate that the direct EOFC approach mitigates the dominant systematic temperature shift.
By measuring the temperature of a \(^{85}\)Rb vapor, we experimentally confirm that direct EOFC Doppler thermometry does not suffer from systematic temperature shifts arising from optical pumping as atoms transit the probe laser beam.
A direct EOFC Doppler thermometer can therefore use a high intensity probe beam to increase signal-to-noise ratio -- and so reduce averaging time -- while maintaining temperature measurement accuracy.
Our work opens a new path toward fast, portable, primary thermometers based on Doppler spectroscopy of atomic vapors.

\begin{backmatter}
\bmsection{Funding}
Our work is funded by the National Institute of Standards and Technology.

\bmsection{Acknowledgments}
We thank J. Lawall and J. Hendricks for their careful reading of the manuscript.
We also thank E. Norrgard for the loan of one of the lasers used in this work.

\bmsection{Disclosures}
S. M. B., S. P. E., D. A. L., B. J. R., and D. S. B. have filed US provisional patent application 24-059P1 (P).

\bmsection{Data Availability Statement}
Data underlying the results presented in this paper are not publicly available at this time, but may be obtained from the authors upon reasonable request.


\end{backmatter}

\bibliography{CombDoppler}

\end{document}